\documentstyle[preprint,aps]{revtex}
\tightenlines

\newtheorem{theorem}{Theorem}

\newtheorem{conjecture}[theorem]{Conjecture}

\begin{document}
\title{Failure of a Stability Conjecture in General Relativity}
\author{Ozay Gurtug and Mustafa Halilsoy}
\address{Dept. of Physics, Eastern Mediterranean University, Gazi Magosa, North
Cyprus, Mersin 10, Turkey}
\date{\today}
\maketitle

\begin{abstract}
By employing an exact back-reaction geometry, Helliwell-Konkowski stability
conjecture is shown to fail. This happens when a test null dust is inserted
to the interaction region of cross-polarized Bell-Szekeres spacetime.
\end{abstract}

\section{Introduction}

Some time ago Helliwell and Konkowski (HK) [1], developed a Cauchy horizon
(CH) stability conjecture that uses test fields to predict the instability
and type of the singularity that forms. Then a complete non-linear back
reaction calculation would show that this type of singularity occurs. The CH
stability conjecture due to HK is defined as follows.

\begin{conjecture}
For all maximally extended spacetimes with CH, the backreaction due to a
field (whose test stress-energy tensor is $T_{\mu \nu }$) will affect the
horizon in one of the following manners. a)If $T_{\mu }^{\mu }$, $T_{\mu \nu
}T^{\mu \nu }$ and any null dust density $\rho $ are finite, and if the
stress energy tensor $T_{ab}$ in all parallel propagated orthonormal (PPON)
frames is finite, then the CH remains non-singular. b)If $T_{\mu }^{\mu }$, $%
T_{\mu \nu }T^{\mu \nu }$ and any null dust density $\rho $ are finite, but $%
T_{ab}$ diverges in some PPON frames, then a non scalar curvature (NSC)
singularity will be formed at the CH. c)If $T_{\mu }^{\mu }$, $T_{\mu \nu
}T^{\mu \nu }$ and any null dust density $\rho $ diverges, then an scalar
curvature (SC) singularity will be formed at the CH.
\end{conjecture}

In ref.[2] (and references therein) the stability conjecture has been tested
for several spacetimes. Among others the exceptional case occurs in the
Bell-Szekeres (BS) [3] spacetime which describes the collision of two
constant profile electromagnetic (em) waves that results in a non-singular
interaction region. It was shown that, when an impacting test em field is
added to one of the incoming regions of the BS spacetime, the conjecture
predicts a non scalar curvature singularity (NSCS). However, the exact
solution due to BS possesses a CH in the region of interaction and
quasiregular singularities at the null boundaries. To our knowledge, this
was a single example shown by HK to signal the failure of this conjecture.%
\newline
In this paper, we further test the validity of this conjecture in the
interaction region of colliding plane wave (CPW) spacetimes admitting two
hypersurface non-orthogonal Killing vectors. For this purpose, we use the
cross-polarized BS (CPBS) [4] solution and the outcome of the conjecture is
compared with an exact back reaction solution. Our analysis verifies the
failure of the conjecture in predicting the type of the singularity for this
case too. \newline
The paper is organized as follows. In section II, we review the CPBS
solution and as a requirement of the conjecture we insert oppositely moving
test null dust in the interaction region of CPBS spacetime. In section III,
we present an exact back-reaction geometry that represents the collision of
null shells (or impulsive dusts) coupled with CPBS spacetime. The paper is
concluded with a conclusion in section IV.

\section{Test Null-Dust in the Space of CPBS Spacetime}

The metric that describes collision of em waves with the cross polarization
was found to be [4] \newline
\begin{equation}
ds^{2}=F\left(\frac{d\tau^{2}}{\Delta}-\frac{d\sigma^{2}}{\delta}%
\right)-\Delta F dy^{2}-\frac{\delta}{F}(dx-q\tau dy)^{2}
\end{equation}
In this representation of the metric our notations are
\begin{eqnarray}
\tau &=& \sin(au + bv)  \nonumber \\
\sigma &=& \sin(au - bv)  \nonumber \\
\Delta &=& 1- \tau^{2}  \nonumber \\
\delta &=& 1- \sigma^{2}  \nonumber \\
2F &=& \sqrt{1+q^{2}}(1+ \sigma^{2}) + 1 - \sigma^{2}
\end{eqnarray}
in which $0 \leq q \leq 1$ is a constant measuring the second polarization, $%
(a,b)$ are constants of energy and $(u,v)$ stand for the usual null
coordinates. It can be seen easily that for $q=0$ the metric reduces to BS.
Unlike the BS metric ,however, this is conformally non-flat for $( u>0, v>0)$%
, where the conformal curvature is generated by the cross polarization. As a
matter of fact this solution is a minimal extension of the BS metric. A
completely different generalization of the BS solution with second
polarization was given by Chandrasekhar-Xanthopoulos [5]. Their solution,
however, employs an Ehlers transformation and involves two essential
parameters which is therefore different from ours. Both solutions form CH in
the interaction region.\newline
Our main interest here is to test HK stability conjecture in the CPBS
spacetimes. In doing this, we insert oppositely moving test null dusts in
the CPBS spacetime. For simplicity we consider two different cases, the $%
x=constant $ and $y=constant $ projections of the spacetime.\newline
We have in the first case ($x=constant $),
\begin{equation}
ds^{2}=\frac{e^{-M}}{2ab}\left(dt^{2}-dz^{2}\right)-e^{-U-V}coshW \,dy^{2}
\end{equation}
where we have used the coordinates $(t,z)$ according to
\begin{eqnarray}
t&=&au+bv  \nonumber \\
z&=&au-bv
\end{eqnarray}
The energy-momentum tensor for two oppositely moving null dusts can be
chosen as
\begin{equation}
T_{\mu \nu}=\rho_{l}l_{\mu}l_{\nu}+\rho_{n}n_{\mu}n_{\nu}
\end{equation}
where $\rho_{l}$ and $\rho_{n}$ are the finite energy densities of the
dusts. The null propagation directions $l_{\mu}$ and $n_{\mu}$ are
\begin{eqnarray*}
l_{\mu}&=&\left(a_{0},0,a_{2},a_{3}\right) \\
n_{\mu}&=&\left(-a_{0},0,a_{2},a_{3}\right)
\end{eqnarray*}
with
\[
a_{2}=k_{2}=\mbox{constant} \hspace{3cm} a_{3}=\frac{k_{1}}{2ab}= %
\mbox{constant}
\]
\[
a_{0}=\frac{1}{2ab}\left(k^{2}_{1}+\frac{2abk^{2}_{2}}{coshW}\,
e^{U+V-M}\right)^{1/2}
\]
We observe from (1) that
\begin{equation}
\frac{e^{U+V}}{coshW} = \frac{F}{\Delta F^{2} + \delta q^{2} \tau^{2}}
\end{equation}
which is finite for $q \neq 0$. Following the requirement of the conjecture,
we find the trace and scalar of the energy-momentum as
\begin{eqnarray}
T^{\mu}_{\mu}&=&0  \nonumber \\
T_{\mu \nu}T^{\mu \nu}&=&2\rho_{l} \rho_{n} \left(\frac{k^{2}_{1} e^{M}}{ab}
+\frac{2k^{2}_{2}e^{U+V}}{coshW}\right)^{2}=\mbox{finite}
\end{eqnarray}
The scalar $T_{\mu \nu}T^{\mu \nu} $ reveals that as $\tau \rightarrow 1 $
it does not diverge and remains finite. Note that for a linear polarization
(i.e. $q=0 $ ), the scalar $T_{\mu \nu}T^{\mu \nu} $ diverges as the horizon
is approached indicating SCS. This particular case overlaps with the work of
HK in ref. [2]. In our case, we use a non-linearly polarized metric and the
outcome conflicts with the result of HK. \newline
Next we consider the energy-momentum tensor in the parallel-propagated
orthonormal frame (PPON). Such frame vectors are
\begin{eqnarray}
e^{\mu}_{(0)}&=& \left( \sqrt{2ab}e^{M/2},0,0,0 \right)  \nonumber \\
e^{\mu}_{(1)}&=& \left( 0,e^{\frac{U-V}{2}}\cosh{W/2},e^{\frac{U+V}{2}}\sinh{%
W/2},0 \right)  \nonumber \\
e^{\mu}_{(1)}&=& \left( 0,e^{\frac{U-V}{2}}\sinh{W/2},e^{\frac{U+V}{2}}\cosh{%
W/2},0 \right)  \nonumber \\
e^{\mu}_{(3)}&=& \left( 0,0,0,\sqrt{2ab}e^{M/2} \right)
\end{eqnarray}
and the energy momentum tensor in this frame is given by
\begin{equation}
T_{(ab)}=e^{\mu}_{(a)}e^{\nu}_{(b)}T_{\mu \nu}
\end{equation}
Then the non-zero components of $T_{(ab)} $ are;
\begin{eqnarray}
T_{00}&=&\left[ \frac{k^2_{2} e^{U+V}}{\cosh W} + \frac{k^2_{1} e^{M}}{2ab}%
\right] \left(\rho_{l}+\rho_{n}\right)  \nonumber \\
T_{01}&=&T_{10}=\left \{ k^2_{2}\left(\cosh W -1 \right)\left[ \frac{
k^2_{2}e^{2(U+V)}}{2 \cosh W }+ \frac{ k^2_{1} e^{U+V+M}}{4ab} \right]
\right \}^{1/2} \left(\rho_{l}-\rho_{n}\right)  \nonumber \\
T_{02}&=&T_{20}=\left \{ k^2_{2}\left(\cosh W +1 \right)\left[ \frac{
k^2_{2}e^{2(U+V)}}{2 \cosh W }+ \frac{ k^2_{1} e^{U+V+M}}{4ab} \right]
\right \}^{1/2} \left(\rho_{l}-\rho_{n}\right)  \nonumber \\
T_{03}&=&T_{30}= \frac{k_{1} e^M}{2ab}\left[ \frac{2ab k^2_{2} e^{U+V-M}}{%
\cosh W} + k^2_{1} \right ]^{1/2} \left(\rho_{l}-\rho_{n}\right)  \nonumber
\\
T_{11}&=&\frac{k^2_{2}e^{U+V}}{2}\left( \cosh W -1 \right)\left(
\rho_{l}+\rho_{n}\right)  \nonumber \\
T_{13}&=&T_{31}=\frac{k_{1}k_{2}}{\sqrt{2ab}}e^{(U+V+M)/2} \sinh
W/2\left(\rho_{l}+\rho_{n}\right)  \nonumber \\
T_{23}&=&T_{32}=\frac{k_{1}k_{2}}{\sqrt{2ab}}e^{(U+V+M)/2} \cosh
W/2\left(\rho_{l}-\rho_{n}\right)  \nonumber \\
T_{33}&=&\frac{k^2_{1}e^M}{2ab} \left(\rho_{l}+\rho_{n}\right)
\end{eqnarray}
Careful analysis of these components shows that some components like ( $%
T_{01},T_{02},T_{11},T_{13},T_{23} $ ) diverge as $\tau \rightarrow 1 $ .
Consequently the conjecture predicts that the CH is unstable and transforms
into an NSCS. \newline
By a similar procedure we investigate the $y= constant $ projection of the
spacetime and obtain that none of the energy- momentum scalars diverges in
the interaction region.\newline

\section{The Geometry of Null Shells with CPW.}

In this section we present an exact back-reaction solution of two colliding
null shells ( or impulsive dusts) coupled with the CPBS spacetime. The
combined metric of CPBS and colliding shells can be represented by[6]
\begin{equation}
ds^{2}=\frac{1}{\phi^{2}}\,ds^{2}_{CPBS}
\end{equation}
where $\phi=1+\alpha u \theta(u) + \beta v \theta(v) $ with $(\alpha, \beta)
$ positive constants. This amounts to the substitutions
\begin{eqnarray}
M&=&M_{0}+2\ln \phi  \nonumber \\
U&=&U_{0}+2\ln \phi  \nonumber \\
V&=&V_{0}  \nonumber \\
W&=&W_{0}
\end{eqnarray}
where $\left(M_{0},U_{0},V_{0},W_{0}\right)$ correspond to the metric
functions of the CPBS solution. Under these substitutions the scale
invariant Weyl scalars remain invariant ( or at most multiplied by a
conformal factor ) because $M-U=M_{0}-U_{0}$ is the combination that arise
in those scalars. The scalar curvature, however, which was zero in the case
of CPBS now arises as nonzero. The Weyl and Maxwell scalars of the new
solution are given in the Appendix.\newline
It is clearly seen that, the scalar curvature diverges as $\tau \rightarrow
1 $ ( or equivalently $au + bv \rightarrow \pi /2 $). It is further seen by
choosing $\beta =0 $, that even a single shell gives rise to a divergent
back reaction by the spacetime. The horizon, in effect, is unstable and
transforms into a SCS in the presence of colliding shells or even a single
propagating null shell.\newline

\section{Conclusion}

In this paper, we have tested the HK stability conjecture in the CPBS
spacetime. Similar analysis was done by HK for the BS spacetime (linear
polarization). However they were unable to compare their results with an
exact back-reaction solution. We confirm their results by taking the limit
as $q \rightarrow 0 $. The line element (1) reduces to the BS and the
expression (6) that appears in equation (7) diverges on the horizon ( $\tau
= 1$) indicating an SCS. For a linearly polarized metric (which is BS ) the
conjecture finds correctly the nature of the singularity. But for
non-colinear polarization (which is the CPBS ) the conjecture predicts a
NSCS. whereas the exact back-reaction solution indicates SCS. Therefore, the
HK stability conjecture fails to predict the correct nature of the
singularity in the non-colinear metrics. Recently, we have also shown [7]
that, the inner horizon of Kerr-Newman black hole, displays a double
character with respect to different perturbing potentials. In the case of
null dust we have shown that the inner horizon is transformed into a
spacelike SCS, however, the inclusion of particular scalar fields creates
null singularities on the inner horizon of Kerr-Newman black hole. All of
these outcomes are supported with exact back-reaction solutions.\newline
Our overall impression about the conjecture is that it can be used to check
the instability of the CHs, but is not reliable in determining the type of
the singularity.\newline

\section*{ Appendix:\newline
The Weyl and Maxwell Scalars}

The non-zero Weyl and Maxwell scalars for the collision of null shells in
the background of CPBS spacetime are found as follows.
\begin{eqnarray}
\Psi_{2}&=& (\Psi_{2})_{(CPBS)}  \nonumber \\
& & \\
\Psi_{4}&=& (\Psi_{4})_{(CPBS)}  \nonumber \\
& & \\
\Psi_{0}&=& (\Psi_{0 })_{(CPBS)}  \nonumber \\
& & \\
4\phi e^{-M}\Phi_{11}&=& \left[ (a\beta + \alpha b)\tan(au+bv) \right.
\nonumber \\
& &  \nonumber \\
& & \left.+(a \beta-\alpha b)\tan(au-bv) \right] \theta(u) \theta(v) \\
& &  \nonumber \\
4\phi e^{-M}\Lambda&=& \left[ (a\beta + \alpha b)\tan(au+bv)+(a \beta
-\alpha b)\tan(au-bv)\right.  \nonumber \\
& &  \nonumber \\
& & \left.+\frac{4\alpha \beta}{\phi} \right] \theta(u) \theta(v) \\
& &  \nonumber \\
\Phi_{22}&=&(\Phi_{22})_{CPBS} + \left(\frac{\alpha e^M }{\phi}\right)\left[
\delta(u) \right.  \nonumber \\
& &  \nonumber \\
& &\left. - \theta(u)\left(a \Pi +\frac{u}{(1-u^2)(1-v^2)}\right) \right] \\
& &  \nonumber \\
& &  \nonumber \\
\Phi_{00}&=&(\Phi_{00})_{CPBS} + \left(\frac{\beta e^M }{\phi}\right)\left[
\delta(v) \right.  \nonumber \\
& &  \nonumber \\
& & \left.+ \theta(v)\left(b \Pi - \frac{v}{(1-u^2)(1-v^2)}\right)\right] \\
& &  \nonumber \\
\Phi_{02}&=& (\Phi_{02})_{CPBS}+\left( \frac{e^M}{4FY\phi}\right) \left[%
\frac{1}{F}\left(\alpha Q \theta(u) + \beta P \theta(v)\right) \right.
\nonumber \\
& &  \nonumber \\
& & \left. +iq\left(\alpha L \theta(u) + \beta K \theta(v)\right)\right]
\end{eqnarray}
where
\begin{eqnarray*}
\phi&=&1+\alpha u \theta(u) + \beta v \theta (v) \\
Q&=&b\left[2q^2\sin(au+bv)\cos(au-bv)-F^2\left(\tan(au-bv)+\tan(au+bv)%
\right) \right. \\
& & \\
& & \left. -2F\cos(au-bv)\sin(au-bv)\left(\sqrt{1+q^2}-1\right)\right] \\
& & \\
P&=&a\left[2q^2\sin(au+bv)\cos(au-bv)+F^2\left(\tan(au-bv)-\tan(au+bv)%
\right) \right. \\
& & \\
& & \left. +2F\cos(au-bv)\sin(au-bv)\left(\sqrt{1+q^2}-1\right)\right] \\
& & \\
Y&=&\left(1+\frac{q^2}{F^2}\tan(au+bv)\sin(au+bv)\cos(au-bv)\right)^{1/2} \\
& & \\
K&=&\frac{a}{\sqrt{\cos(au+bv)\cos(au-bv)}}\left[\frac{\cos(au-bv)}{
\cos(au+bv)}+\sin2au \right. \\
& & \\
& & \left. -\frac{2\left(\sqrt{1+q^2}-1\right)\sin(au+bv)\cos(au-bv)
\tan(au-bv)}{F}\right] \\
& & \\
L&=&\frac{b}{\sqrt{\cos(au+bv)\cos(au-bv)}}\left[\frac{\cos(au-bv)}{
\cos(au+bv)}+\sin2bv \right. \\
& & \\
& & \left. +\frac{2\left(\sqrt{1+q^2}-1\right)\sin(au+bv)\cos(au-bv)
\tan(au-bv)}{F}\right] \\
& & \\
\Pi&=&\frac{\left(\sqrt{1+q^2}-1\right)\sin(2au-2bv)}{\sqrt{1+q^2}+1+ \left(%
\sqrt{1+q^2}-1\right)\sin^2(au-bv)}
\end{eqnarray*}
and the subscript (CPBS) refers to the expressions given in ref. [4].

\end{document}